# Quantitative schlieren imaging coupled with inverse Abel transformation for flow field studies in a cylindrical surface DBD

K. Giotis[1,2], A. G. Makrypodi[1], P. Svarnas[1]†, K. Gazeli[2], G. Lombardi[2]

[1]*Univeristy of Patras, Electrical and Computer Engineering Department, High Voltage Laboratory, Plasma Technology Group, 26 504 Rion – Patras, Greece*

[2]*University Sorbonne Paris Nord, LSPM, CNRS, 8 UPR 3407, F-93430 Villetaneuse, France*

†*Corresponding author:* svarnas@ece.upatras.gr

**Abstract**

The present work is devoted to the proper application of the schlieren imaging technique, combined with inverse Abel transformation, to record the flow field induced by atmospheric pressure cold plasmas, in a quantitative manner. Theoretical principles, along with meaningful diagrams, design of the schlieren setup, and involved calibration process, are given in detail. It is demonstrated that erroneous conclusions may be reached if Abel transformation is omitted. The concept is employed to study flow field parameters around a surface dielectric barrier discharge, operating in ambient air and having a cylindrical configuration, rather than a typical planar design. Maximum density and lowest temperature equal to 1.24 kg $m^{-3}$ and 285.4 K, respectively, and lowest density and maximum temperature equal to 1.11 kg $m^{-3}$ and 318.8 K, respectively, downstream of the driven electrode and close to the surface are identified with respect to the unperturbed air where density equals 1.21 kg $m^{-3}$ and temperature equals 291.5 K. The temperature and density patterns unveil intense perturbations at shorter distances from the surface and lower amplitudes of the sinusoidal (10 kHz) voltage that sustains the discharge. Electrohydrodynamic effects seem to dominate over thermal mechanisms in governing the flow field. The importance of the quantitative schlieren technique, as a non-invasive one, in the study of dielectric barrier discharges is justified by the interest that they attract in numerous demanding cases of plasma-assisted flow field control (propulsion, actuators, *etc.*).

## 1. Introduction

Atmospheric-pressure plasmas are characterized by unique fluid-mechanical features, enabling their exploitation in diverse aerodynamic applications. Notable examples encompass plasma propulsion and actuators. Thus, plasma-assisted propulsion has been demonstrated in the case of a small aircraft[1], a flying micro-robot[2], as well as in a miniature boat[3], whereas actuators have been applied to the stall improvement of a small unmanned aerial vehicle[4], the airflow optimization of wind turbines[5], *etc*. In such applications, gas density and temperature are key parameters of the flow field, since they may be correlated with the physical mechanisms involved and thus support optimization of the plasma-based systems. Accordingly, either appropriate diagnostics are employed to measure both quantities or relevant theory is used to estimate the one quantity as far as the other one is measurable.

The gas temperature may be directly measured by means of thermal probes, *i.e.*, thermocouples[6–10] or fiber optic thermometers. Thermocouple operation is in principle based on Seebeck effect, whereas fiber optic thermometer operation may be based either on a Bragg grating instrument[6,7] or the thermal expansion of a highly stable glass[11] or the dependence of the bandgap of a semiconductive crystal on the temperature[12–14] or the dependence of the fluorescent signal from a luminescent material on the temperature[15], *etc*. However, independently of the principle of operation, such approaches may be intrusive for a plasma, especially in the case of conductive probes. On the other hand, spectroscopic techniques are employed widely. A common method refers to the recording of the rotational distribution of diatomic molecules (*e.g.*, $N_2$, NO, OH), since rotational temperature may be justified as a fair approximation of the translation (gas) temperature under certain conditions[16]. In view of that, rotational temperature is measured through optical emission spectroscopy[6,9,10,17,18], optical absorbance spectroscopy[19,20], laser induced fluorescence[21,22], and Raman spectroscopy, *i.e.*, either coherent Anti-Stokes Raman spectroscopy[23] or spontaneous Raman scattering[24,25]. Another method to measure the gas temperature involves the observation of the line profile of an emission or absorption transition (usually of an atom)[26–29]. Briefly, any transition is inherently related to spectral width due to different effects, like natural, Stark, Doppler, van der Waals, and resonance broadening, with the last three mechanisms to be dependent on the gas temperature[30].

Concerning the gas density, it may be also derived from optical diagnostics. For instance, Rayleigh scattering is used[31–33], which corresponds to the elastic scattering of photons on electrons bound to neutrals. The scattering cross-section depends on the species, while the intensity of the scattered light is proportional to gas density[34]. In addition, since the refractive index of a gas is related to its density, interferometric methods are employed, like Mach-Zehnder interferometry[35–37], which follows the phase difference between a reference and a probing light ray. Then, it is known that any change in the refractive index of a gas medium leads to the bending of any light ray passing through it[38]. The quantification of such deflections can be used to determine the refractive index, which in turn enables the evaluation of the gas density either by means of shadowgraphy[39] or schlieren imaging[40]; shadowgraphs display light ray displacement resulting from the deflection angle whereas schlieren images display light ray deflection angle.

Especially schlieren imaging has been extensively used for qualitative characterization of the flow field induced by atmospheric-pressure cold plasmas. Reports refer to plasma jets[41–45], surface dielectric barrier discharge (SDBD) actuators[46–50], and sparks[51,52]. Nevertheless, the quantification of the corresponding results is discussed in a disproportionally extend in the literature. In those works: the gas temperature is evaluated in a DBD argon plasma jet[6] and an air DC corona discharge[53], by means of Z-type schlieren imaging; the gas density is obtained in air SDBDs under AC[54–56] and ns-pulsed[57] excitation, using a background oriented schlieren and a Z-type schlieren configuration, respectively; both gas and temperature are estimated in an RF argon plasma jet[58] and an impulsive air discharge (leader regime)[59], with a Toepler-type schlieren setup; and, finally, both gas and temperature are assessed in an air arc[60] and in a low-frequency oxygen plasma jet[61], using a Z-type configuration.

As a contribution to this scientific field, the present work comprises a reference to the proper application of quantitative schlieren imaging, coupled with inverse Abel transformation, to probe both the gas temperature and density in flow fields induced by atmospheric-pressure cold plasmas. The fundamental principles along with detailed diagrams, the design of the experimental setup employed, and the involved calibration process, are here detailed and ultimately established during the study of a special electrical discharge. This refers to a cylindrical SDBD produced by a configuration which tackles numerous technical issues confronted during the operation of conventional SDBDs, as it has been reported previously[62,63]. Furthermore, the present study comes to fill an important gap in literature where numerous studies on planar SDBDs do exist, but quantitative considerations of curved plasma-air interfaces by means of quantitative schlieren technique are scarce. Finally, contrary to most prevailing reports, wide domains are here probed. Thus, plasma-induced flow field patterns transit to the surrounding unperturbed air are visualized.

## 2. Experimental setup

The general view of the experimental setup is depicted in **Fig. 1a**. The entire system is installed on an optical table (Thorlabs Inc.; T1020C; Nexus; 1 m × 2 m × 0.21 m). The plasma reactor, shown in **Fig. 1b** along with the corresponding coordinate system, is designed to produce a cylindrical SDBD, operating in quiescent ambient air at atmospheric pressure. It comprises an arrangement of two stainless-steel electrodes. The outer (driven) one is shaped as a hollow truncated cone and the inner (grounded) one forms a cylindrical tube. These electrodes are separated by a quartz dielectric tube, filled with transformer oil, while the vertical gap between their edges is set to zero. More technical details and insights into the discharge dynamics and the emission characteristics, are given in previous reports[62,63].

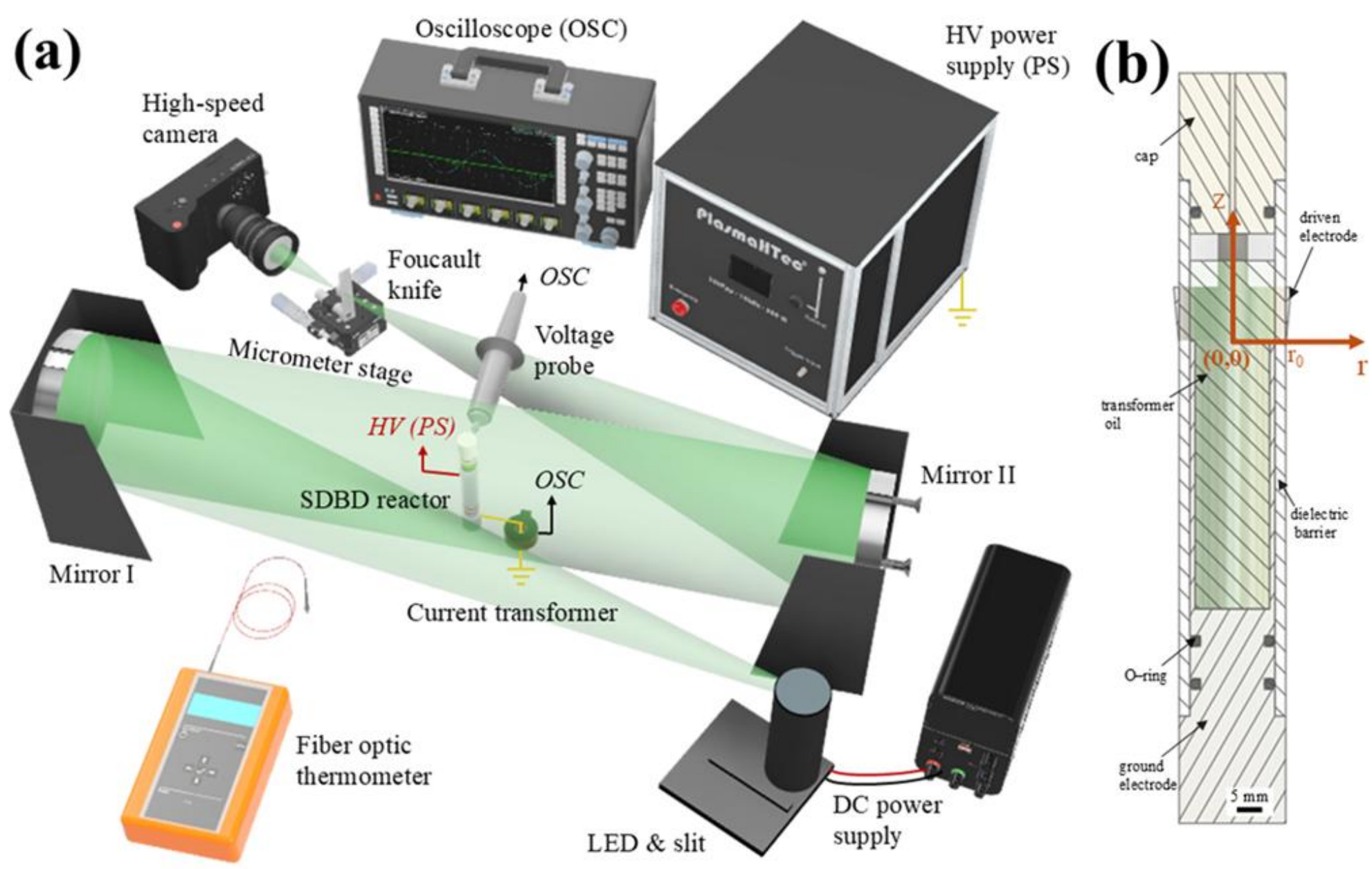

**Figure 1. (a)** General view of the experimental setup built around the cylindrical SDBD reactor shown at the center. **(b)** Longitudinal cross section of the cylindrical SDBD reactor filled with transformer oil of high dielectric strength. The reader is referred to the text and the cited references for further description.

The reactor is driven by a sinusoidal AC high voltage power supply (PlasmaHTec®). Its specifications include voltage amplitudes up to 15 kV (peak), real power up to 500 W, and fixed frequency of 10 kHz with a total harmonic distortion down to 1%. This power supply is a fully standalone device and thus any external auxiliary unit is not needed for its operation. The driving voltage and the induced current are monitored through a high voltage probe (Tektronix; P6015A; DC – 75 MHz) and a current transformer (Pearson Electronics; 6585; 400 Hz – 200 MHz), respectively, which are matched to a digital oscilloscope (Teledyne LeCroy; WaveRunner 44Xi-A; 400 MHz; 5 GSamples $s^{-1}$ per channel). Indicative oscillograms of the driving voltage and induced total current are provided in **Fig. 2a**, while the physical interpretation of these correlated waveforms can be found in previous report[62]. Furthermore, the visible aspect of the discharge propagating on the reactor surface with increasing voltage is shown in **Fig. 2b** (time-integrated photos).

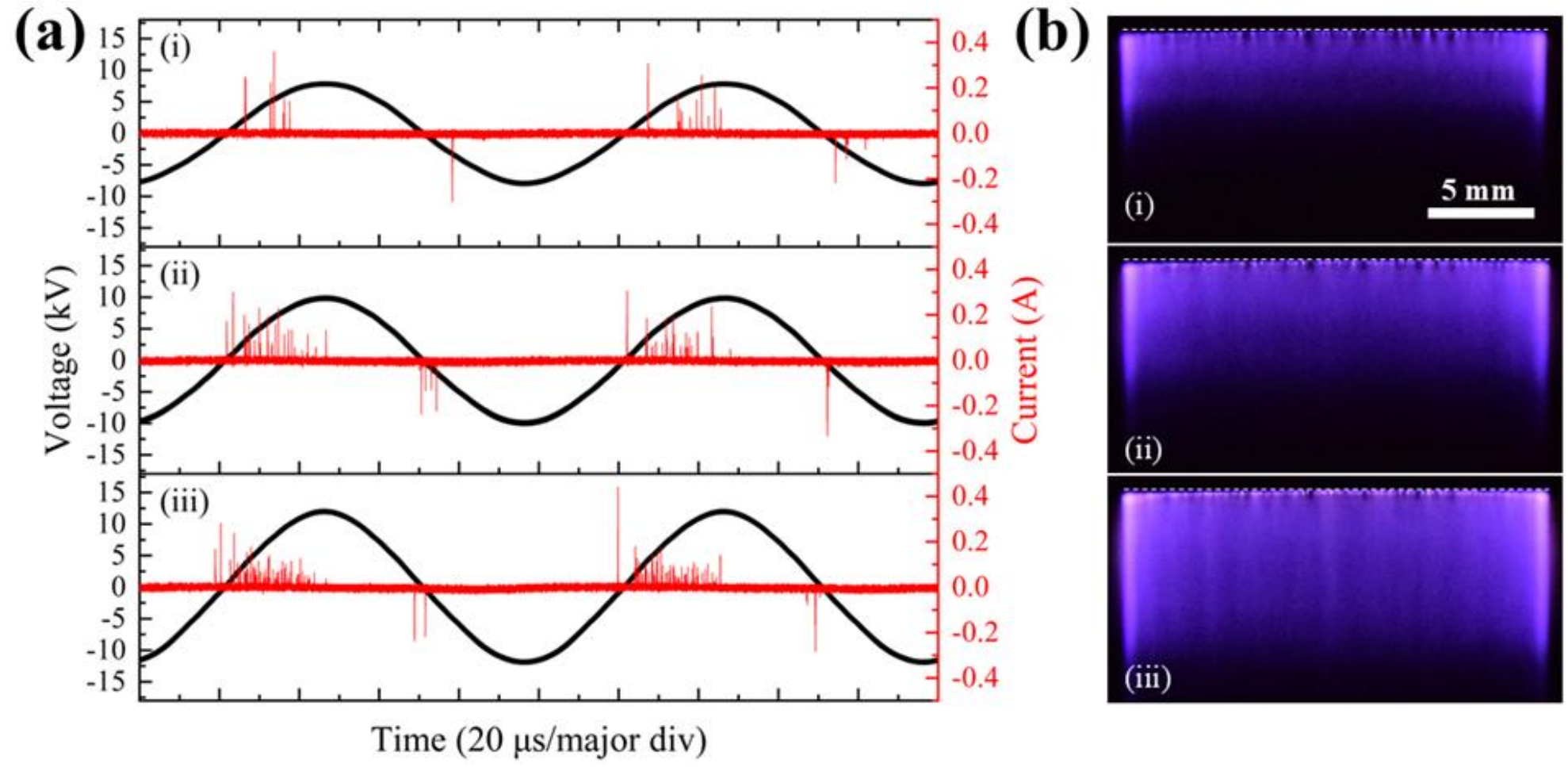

**Figure 2. (a)** Typical waveforms of the driving voltage and the induced current of the plasma reactor, while the voltage amplitude increases; **(i)** 16 $kV_{pp}$, **(ii)** 20 $kV_{pp}$, **(iii)** 24 $kV_{pp}$. **(b)** Visual inspection (conventional photos) of the discharge sustained along the quartz surface for increasing voltage amplitude; **(i)** 16 $kV_{pp}$, **(ii)** 20 $kV_{pp}$, **(iii)** 24 $kV_{pp}$.

A schlieren imaging system, **Fig. 1a**, is installed to visualize the flow field in the air surrounding the plasma reactor and, eventually, determine the gas density and temperature in a spatially resolved manner. The system is arranged in Z-type configuration, which includes a high-power green light-emitting diode (*i.e.*, LED; Luminus; PT-121-TE; 525 nm central peak; 34 nm FHWM; 4×3 $mm^2$), a circular diaphragm (Ø5 mm), two spherical mirrors (Edmund Optics®; Ø152.4 mm; 1524 mm focal length; λ/8 surface accuracy), a razor (Foucault knife), and a high-speed camera (Kron Technologies; Chronos 2.1-HD; 1920×1080 pixels maximum resolution; 24046 fps maximum; 12 bit depth; 10 μm pixel pitch) equipped with an appropriate objective lens (Tamron; 28-300 mm focal length; f/3.5-6.3 maximum aperture; 10.7 max. zoom ratio). The LED and the Foucault knife are mounted vertically at the focal point of each mirror, respectively (**Fig. 1a**).

The Foucault knife is placed on a 2-axis micrometer translation stage for fine adjustments. It blocks part of the projected light for displacements along the horizontal, "left-right" direction. On the other hand, following a fine adjustment along the "back-forward" direction, an intermediate position where the camera image is illuminated as uniformly as possible is found; thereafter the position at this "back-forward" direction does not change. As a measure of uniformity, the standard deviation of the intensity of the sensor pixels corresponding to the green light is numerically calculated and values lower than 6% are considered acceptable.

It is worthy to note that at this optimal "back-forward" position, any short displacement of the knife along the "left-right" direction produces uniformly illuminated camera images (of course when plasma is OFF).

In addition, the LED is driven by a DC power supply (NANKADF®; 150 W) operating in constant-current mode (here 5 A), while 1-h stabilization period of the LED is done before each experimental set. The light emitted by the LED passes through the circular diaphragm and propagates in a conical form. The rays are collimated by the first mirror (Mirror I in **Fig. 1a**) and subsequently are focused onto the Foucault knife by the second mirror (Mirror II in **Fig. 1a**). The focus plane is the vertical one (parallel to both mirrors) which passes from the reactor symmetry-axis and the resulting depth of field has been measured to be about ±50 mm behind/in front of this plane. The camera operates in internal mode (ISO 500), and the recorded images consist of 800 ×1080 pixels corresponding to 67.2 × 90.72 mm$^2$ physical area. The frame rate and the exposure time are set to 2000 fps and 100 μs, respectively. All snapshots are captured after a 20-min warming up period, ensuring thermal stabilization of all the material comprising the plasma reactor. Lastly, a fiber optic thermometer (LumaSense Technologies Inc.) is used to determine the refraction index of the surrounding undisturbed air (as it will be explained below).

Conventional photos of the plasma emission (corresponding mainly to the visible range of the spectrum; **Fig. 2b**) are captured by a digital single lens reflex (DSLR) camera (Nikon D3300; 1/50 s exposure time; ISO-6400) equipped with an objective lens (Sigma; focal length 24–70mm; maximum aperture f/2.8).

## 3. Fundamental principles

Schlieren images can be utilized for quantitative characterization of the flow field based on the following theoretical principles. When light passes through an inhomogeneous medium, the rays deviate by an angle (**Fig. 3a**). This is due to the spatial variation of the refractive index. Specifically, the density and refractive index $n$ of a transparent medium are governed by the Lorenz-Lorentz formula[64]

$$\frac{n^2-1}{n^2+2} = \frac{4\pi}{3}\sum_i N_i \alpha_i, \tag{1}$$

where $N_i$ is the molecule density and $a_i$ is the mean polarizability of species $i$. In any gas, the refractive index is close to 1, $n \cong 1$, and consequently the formula results to the empirical Gladstone-Dale equation[65]

$$\frac{n-1}{\rho} = k, \tag{2}$$

where $\rho$ stands for the gas density and $k$ for the Gladstone-Dale constant which depends on the gas composition and light wavelength. At the same time, it is known that under a given set of physical conditions the equation of state correlates the gas temperature ($T$) with the gas density ($\rho$), through the formula

$$T = \frac{p}{R_s \rho}, \tag{3}$$

where $p$ stands for the gas pressure and $R_s$ stands for the specific gas constant. By the geometric theory of the refraction, for a given length ($L$) of propagation of light rays to the direction $y$, the final resulting angles ($\varepsilon_x$ and $\varepsilon_z$ in **Fig. 3a**) of the deviated rays ("deviation" is defined with respect to the initial propagation direction $y$), are related to the refractive index of the medium and the given length, through the equations[38]

$$\varepsilon_x = \int_L \frac{1}{n}\frac{dn}{dx} dy \tag{4}$$

and

$$\varepsilon_z = \int_L \frac{1}{n}\frac{dn}{dz}dy. \tag{5}$$

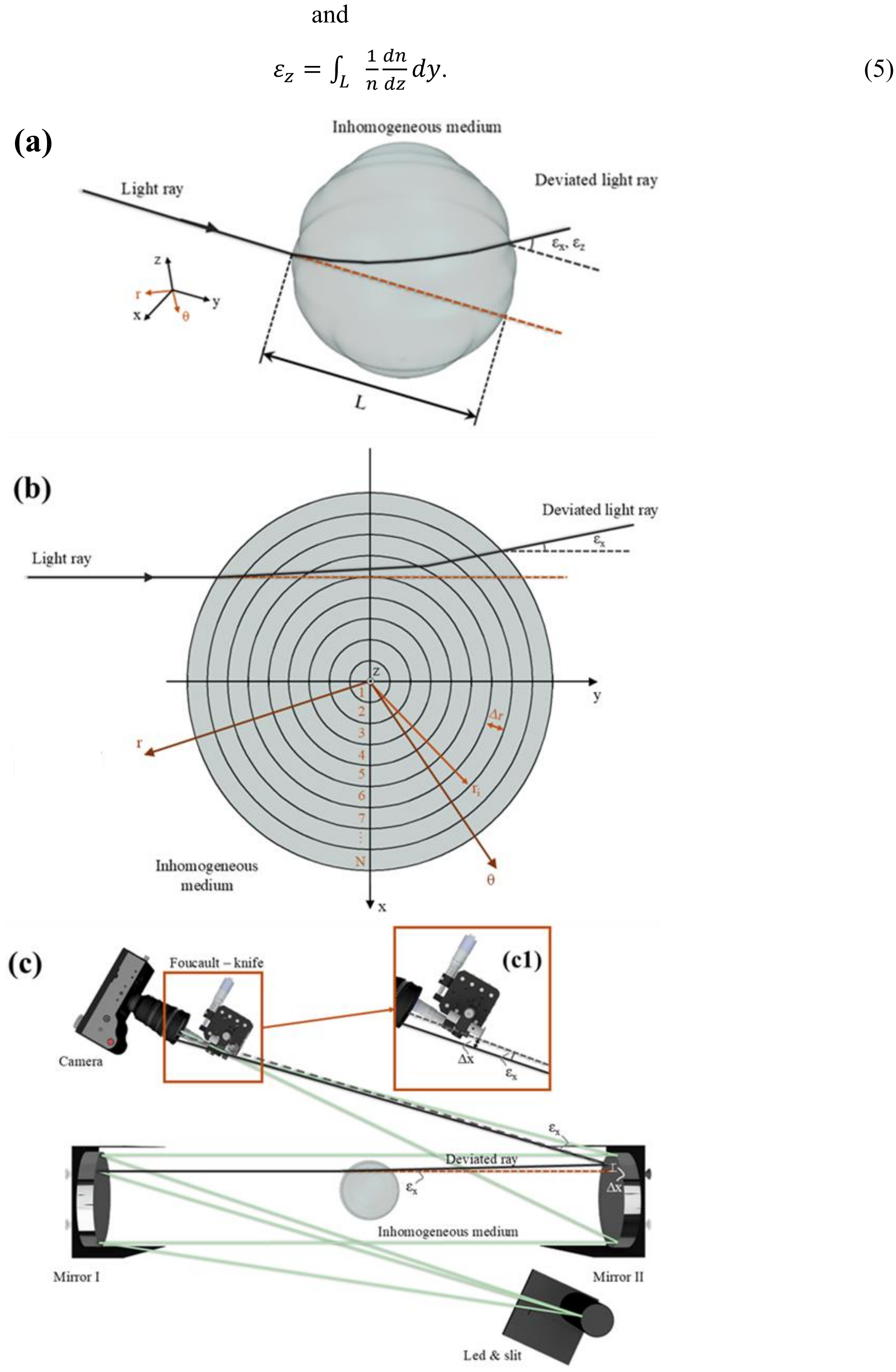

**Figure 3. (a)** Cumulative deviation angle of a light ray traversing an inhomogeneous medium; propagation along the *y*-axis and bending at *x*- and *z*-axis. **(b)** Discretization in the cylindrical coordinate system of the domain of the inhomogeneous medium, to apply the inverse Abel transformation. **(c)** Definition of the deviation angle $\varepsilon_x$ and displacement $\Delta x$ of the light ray in the Z-type schlieren system used to the present study. The inhomogeneous medium corresponds to the (grey) circular disk shown between the two mirrors. Inset **(c1)** zooms in the vicinity of the Foucault knife where the deviated ray (solid line) is presented with respect to the non-deviated case (dash line).

In many practical cases, an axisymmetric refractive index field can be exhibited along one axis at least. Then, using the relative refractive index difference, $\delta = [(n / n_{ref}) - 1]$, where $n_{ref}$ is the refractive index of the surrounding undisturbed-reference medium, the relation becomes (case of $x$-axis deviation angles with axisymmetry around the $z$-axis)[66]

$$\varepsilon_x = 2x \int_x^{\infty} \frac{d\delta}{dr} \frac{dr}{\sqrt{(r^2 - x^2)}} . \tag{6}$$

As the deviation angle is integrated in the line of sight, **Fig. 3b**, the inverse Abel transformation can be applied to provide the radial distribution of the refractive index, *i.e.*:

$$\delta(r) = -\frac{1}{\pi} \int_r^{\infty} \varepsilon_x(x) \frac{dx}{\sqrt{(x^2 - r^2)}} . \tag{7}$$

To solve this integral, it may be split into a sum of integrals, factoring out the deviation angle and approximating the deviation angle by means of linear interpolation[67]. This results in the expression

$$\delta(r_i) = -\frac{1}{2\pi} \sum_{j=i}^{N} [\varepsilon_{x\,j} + \varepsilon_{x\,j+1}] \int_{r_j}^{r_{j+1}} \frac{dx}{\sqrt{(x^2 - r_i^2)}} , \tag{8}$$

where $r_i = i \times \Delta r$ is the radial distance from the centerline, $\Delta r$ is the sampling interval, and $N$ is the total number of intervals in the medium under question (**Fig. 3b**).

Finally, the analytical solution is given as

$$\delta(r_i) = \sum_{j=i}^{N} D_{ij} \varepsilon_{x\,j}, \tag{9}$$

where

$$D_{ij} = \begin{cases} J_{ij}, & j = i \\ J_{ij} + J_{i,j+1}, & j > 1 \end{cases}, \tag{10}$$

with,

$$J_{ij} = -\frac{1}{2\pi} \ln \left[ \frac{j+1+[(j+1)^2 - i^2]^{1/2}}{j + (j^2 - i^2)^{1/2}} \right]. \tag{11}$$

In **Fig. 3c**, the deviation angle and the corresponding displacement are illustrated in a Z-type schlieren configuration, where the Foucault knife blocks the rays horizontally. Under this arrangement, the intensity varies only due to the deviation angles along the $x$-axis (see **Appendix** for justification). Hence, in the case of gas media and relatively weak perturbations, the deviation angles are acceptably small, and a deviation angle is related to a corresponding displacement $\Delta x$ in the Foucault knife plane as follows[38]:

$$\varepsilon_x \approx tan(\varepsilon_x) = f_2 \cdot \Delta x, \tag{12}$$

where $f_2$ is the focal length of mirror II in **Fig. 3c**.

## 4. Experimental method

Based on the above-mentioned fundamentals, the methodology outlined in the flowchart given in **Fig. 4a** is employed to determine the gas density and temperature, using a Python-based implementation. In **Fig. 4b**, representative process inputs and outputs are included in each stage of the method. Being at the optimal "back-forward" position explained in Section 2, initially, the Foucault knife is being displaced along the horizontal,

"left-right", direction to determine the position where the raw images captured with the plasma ON (i) adequately distinguish the features of the induced flow field (increased contrast) and (ii) ensure operation within the non-saturated parts of the calibration curves of the schlieren system (see next paragraph). These two conditions are finally fulfilled following trial and error iterations. Thereafter, the resulting acceptable position remains the same during all experiments with the plasma ON.

On the other hand, the image captured at this position when the plasma turns to OFF is hereafter called "reference schlieren image", **Fig. 4bi**. This reference image is subtracted from any raw one captured with the plasma ON (**Fig. 4bii**). For this subtraction only the pixels of the camera sensor which correspond to the green light are considered. The developing image is given in **Fig. 4biii**. Then, the deviation angle, **Fig. 4biv**, is extracted through a calibration process, as follows.

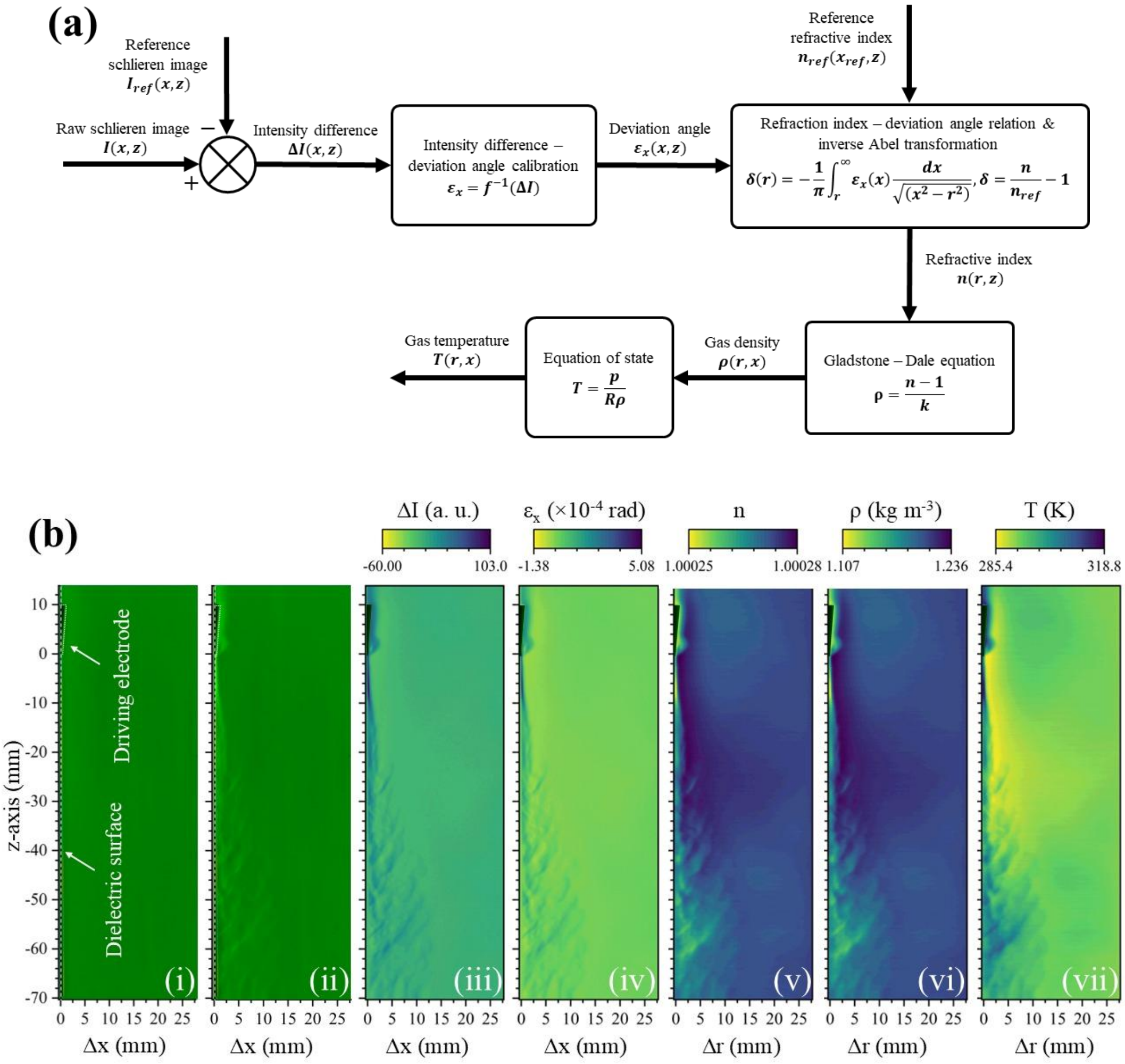

**Figure 4. (a)** Block diagram of the experimental method for quantitative measurements. **(b)** Representative process inputs and outputs in each stage of the method: **(i)** reference schlieren image; **(ii)** raw schlieren image; **(iii)** intensity difference ($\Delta I$); **(iv)** deviation angle ($\varepsilon_x$); **(v)** refractive index ($n$); **(vi)** gas density ($\rho$); **(vii)** gas temperature ($T$). The outer boundary of the dielectric surface is at $\Delta x = 0$ or $\Delta r = 0$ (thin dashed white line marked in (i)), while the driving electrode is the black area in the vicinity of the dielectric surface at $10 \geq z \geq 0$, marked in (i).

The concept of the calibration process is demonstrated with the simplified optical geometry depicted in **Fig. 5**. Light rays propagate along the Z-type schlieren system, beginning from an extended light source and arriving up to the detecting camera. If the light source was an idealized one (point light source), that emits light equally in all directions from a single infinitesimally small point in space, all the rays would be focused on a single infinitesimally small point on the camera side too. There, any displacement of the Foucault knife would lead to an "ON–OFF" situation, *i.e.*, to either a very bright image of uniform intensity (knife does not intersect the single focusing point) or to a fully dark one (knife does intersect the single focusing point)[38].

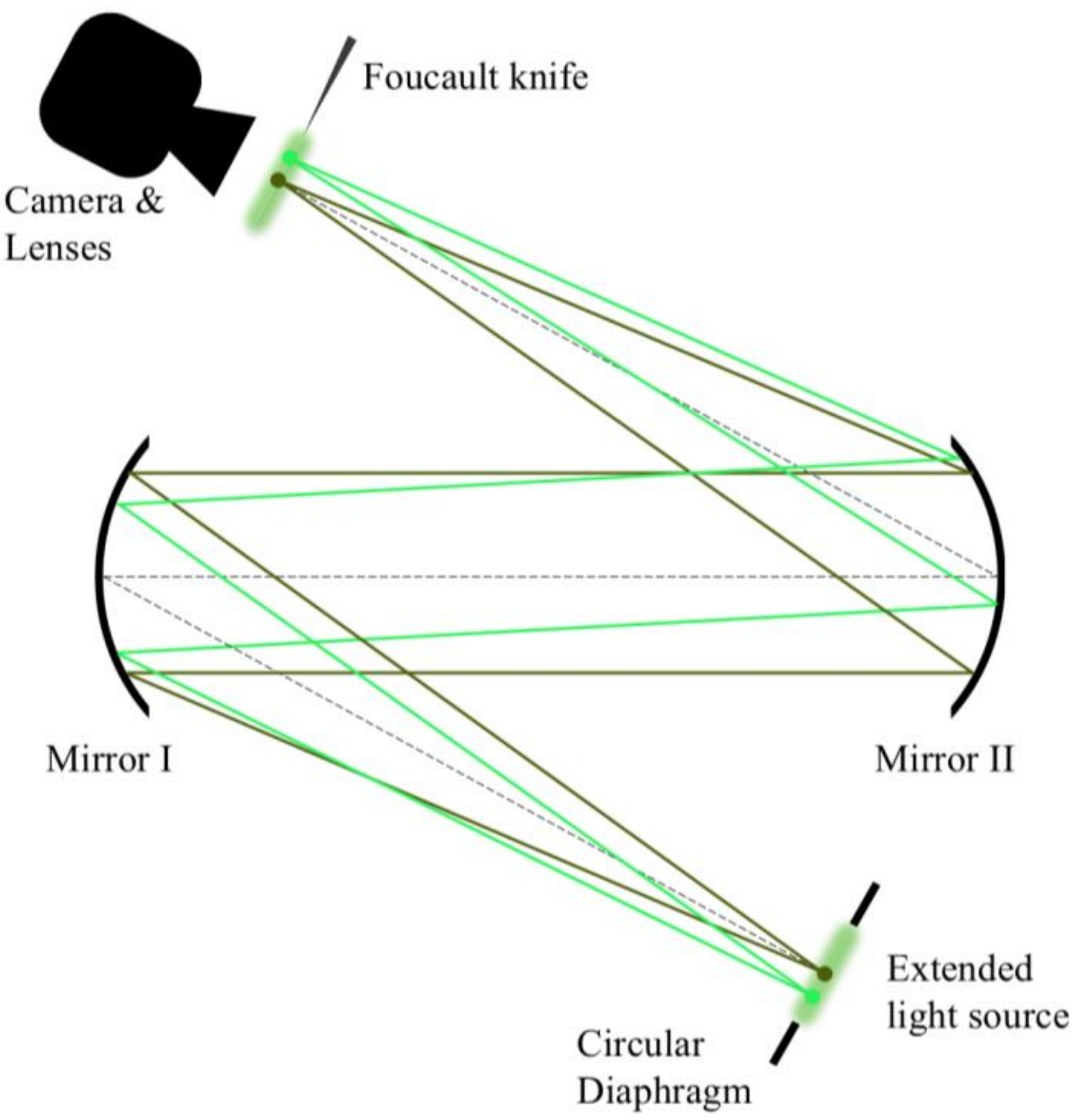

**Figure 5.** Simplified optical geometry, demonstrating light rays generated from an extended source and then propagate through a Z-type schlieren system up to the corresponding detector. Black dotted line: optical axis of the system; Dark green line: rays being parallel to the optical axis (not deviated) between the two mirrors; Bright green line: rays forming an angle with the optical axis (deviated) between the two mirrors.

But, for an extended light source as in practical cases (**Fig. 5**), different aligned emitting points of the light source, lead to correspondingly different aligned focusing points on the camera side. Now, any displacement of the Foucault knife in a way to intersect a specific part of those focusing points, leads to a specific light intensity recorded by the detector (binary states between 0 and 255 in the case of digital detector), spanning from a very bright image (no intersection) down to a fully black image (full intersection). It must be noted that the image produced at any distinct position of the knife is of uniform intensity (*i.e.*, all the sectors on the detector are equally illuminated; deviation < 6%), independently of the intensity's value itself, because of the optimal placement of the knife. Thus, the complexity of "pixel-to-pixel" calibration is avoided[xx].

The above discussion implies that any angle formed between a light ray and the optical axis, between the two mirrors, for any reason, is univocally correlated to a light intensity, assigning specific values (digital states) on the detector pixels (equal value for all pixels). In the present work, these angles are caused by the ray deviations due to the local variation of the diffraction index induced by the plasma-air interaction, and their values (plasma ON) are determined based on the calibration curves (plasma OFF).

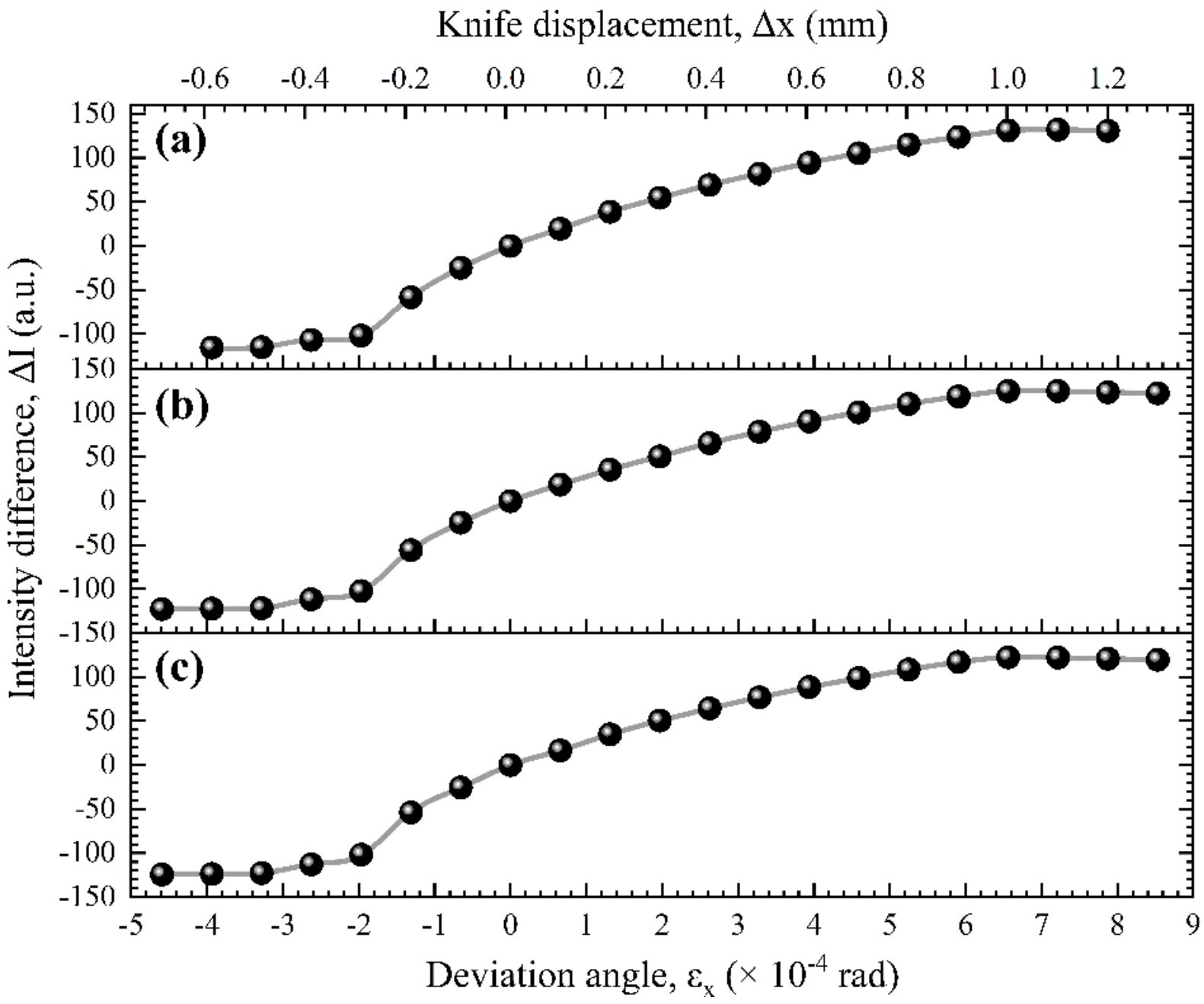

**Figure 6.** Calibration curves, *i.e.*, experimental determination of the relation between light intensity difference and deviation angle of the light rays, for the operating conditions tested: **(a)** 16 $kV_{pp}$; **(b)** 20 $kV_{pp}$; **(c)** 24 $kV_{pp}$.

Accordingly, calibration curves are achieved with the plasma OFF, by horizontally (left-right) displacing the Foucault knife ($\Delta x$) over a wide range and capturing the corresponding images. The intensity difference due to each ray deviation is calculated by the subtraction of the "reference schlieren image" from each other image, while the deviation angles are related to the displacement $\Delta x$ based on **Eq. 12**. The resulting calibration curves in the case of the present work are illustrated in **Fig. 6**. There, quadratic spline interpolation has been applied, numerically, to evaluate intensity difference values corresponding to deviation angles that fall between known experimental data.

Afterwards, the refractive index (plasma ON), **Fig. 4bv**, is extracted by solving the inverse Abel transformation equation (refractive index vs. deviation angle; **Eq. 7**) for each image row. As a "reference refractive index", $n_{ref}$, is taken the one estimated at the farthest points of the image. To determine this value, the temperature is measured by means of the fiber optic thermometer and, subsequently, the equation of state (**Eq. 3**) and the Gladstone-Dale equation (**Eq. 2**) are applied. The following values are considering for that: $p$ = 1 Atm, $R_s$ = 287.086 J kg$^{-1}$ K$^{-1}$, and $k$ = 2.272 × 10$^{-4}$ m$^3$ kg$^{-1}$ at the wavelength of 525 nm[68]. Then, taking the refractive index data (for plasma ON) over the radial and vertical axes, **Eq. 2** is used to calculate the gas density (**Fig. 4bvi**). Once more, the value $k$ = 2.272 × 10$^{-4}$ m$^3$ kg$^{-1}$ is taken at 525 nm[68], under the assumption that the density of the species produced by the plasma are negligible as compared to the ambient air neutral density (*i.e.*, low ionization degree[69]). Finally, the gas temperature (**Fig. 4bvii**) is derived by the equation of state assuming gas pressure $p$ = 1 Atm (due to relative low flow velocities[70]) and using the specific gas constant of air, 287.086 J kg$^{-1}$ K$^{-1}$, for the gaseous phase (same justification made for the constant $k$).

## 5. Results and Discussion

In this section, principal results obtained from the application of the above quantitative method to the plasma produced by the SDBD reactor described in Section 2, are given. The results prove the validity of the method as well as its importance to get access to physical parameters that otherwise would be difficult to record, without any disruption of the gas flow field.

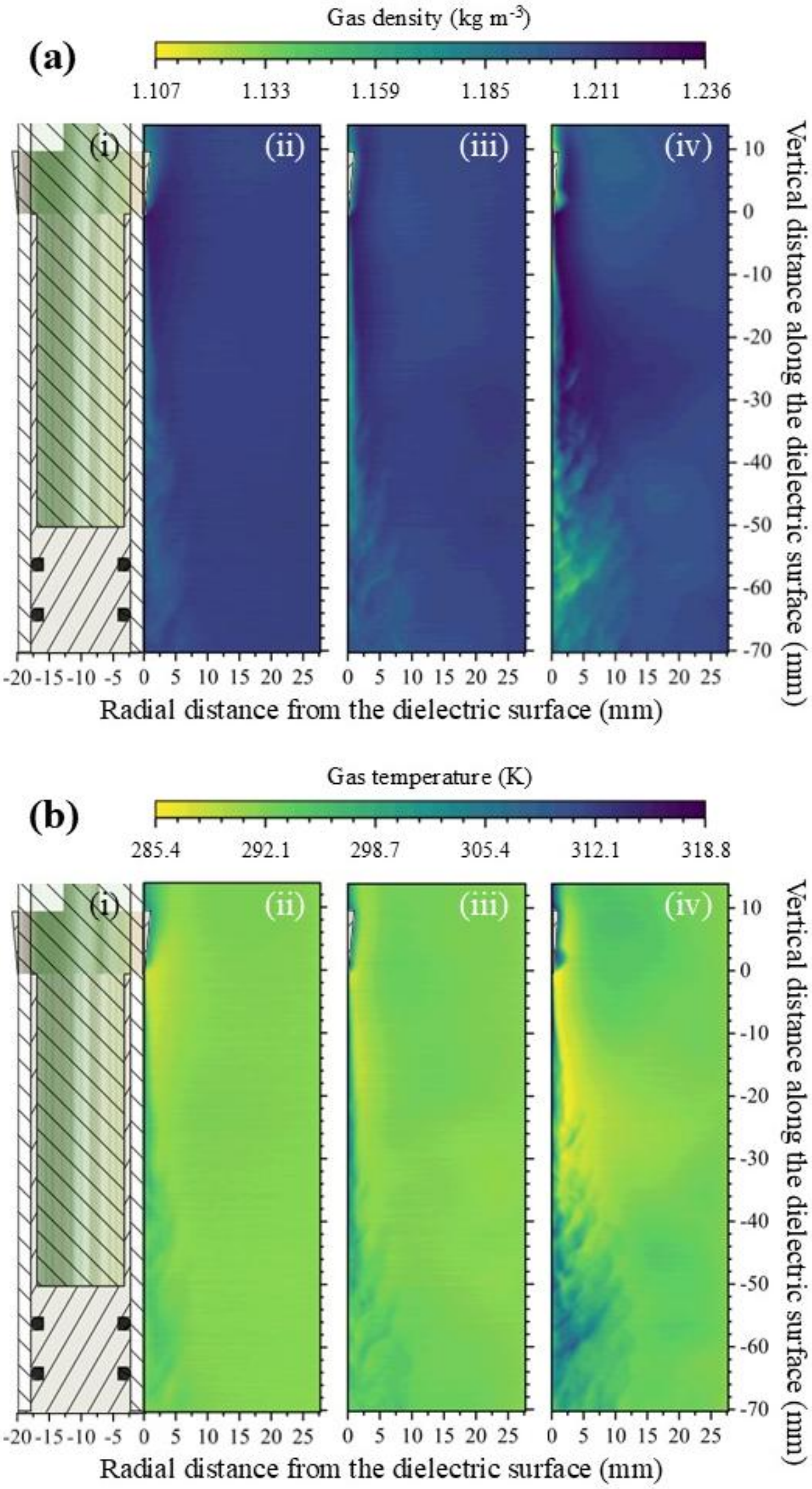

**Figure 7. (ai) & (bi)** Longitudinal section of the plasma reactor. Spatial patterns of the **(a)** gas density and **(b)** temperature, with the operating driving voltage as a parameter: **(ii)** 16 $kV_{pp}$; **(iii)** 20 $kV_{pp}$; **(iv)** 24 $kV_{pp}$. Ambient conditions: 291.5 K and 1.21 kg $m^{-3}$.

**Fig. 7a** and **7b** maps the air density and temperature around the reactor for increasing values of the driving voltage, respectively. As a first remark, air properties vary within distances much longer than the visible pattern of the discharge itself (**Fig. 2b**) and perturbations at least 70 mm downstream of the driven electrode (**Fig. 7**) are observed. This scheme becomes more pronounced as the voltage increases. As a second remark, the perturbated air follows a motif of two distinguishable regions: the first one is attached to the driven electrode and appears as a laminar layer, whereas the second follows this layer and appears as a turbulent extended domain. As the voltage increases from 16 kV to 20 kV (peak-to-peak), from **Fig. 7aii/bii** to **Fig. 7aiii/biii**, the laminar layer is lengthened, the air temperature inside increases, and the air density falls accordingly. At the same time, the domain that follows is expanded and the air inside becomes more perturbated. For further increase of the voltage, **Fig. 7aiv/biv**, this domain expands towards the driven electrode, compromising the laminar layer, becoming highly inflated and perturbated. Eddy-type regions in this domain are now associated with elevated air temperatures and lowered air densities. As a third remark, the flow field and the temperature/density of the air adjacent the driven electrode, are also modified. Thus, thermal and electro-hydrodynamic effects take place within the vicinity of both the dielectric and the conductive components of the reactor. Sucking effect is probably exhibited close to the driven electrode, as is argued by the low density found there (**Fig. 7aiii/biii**).

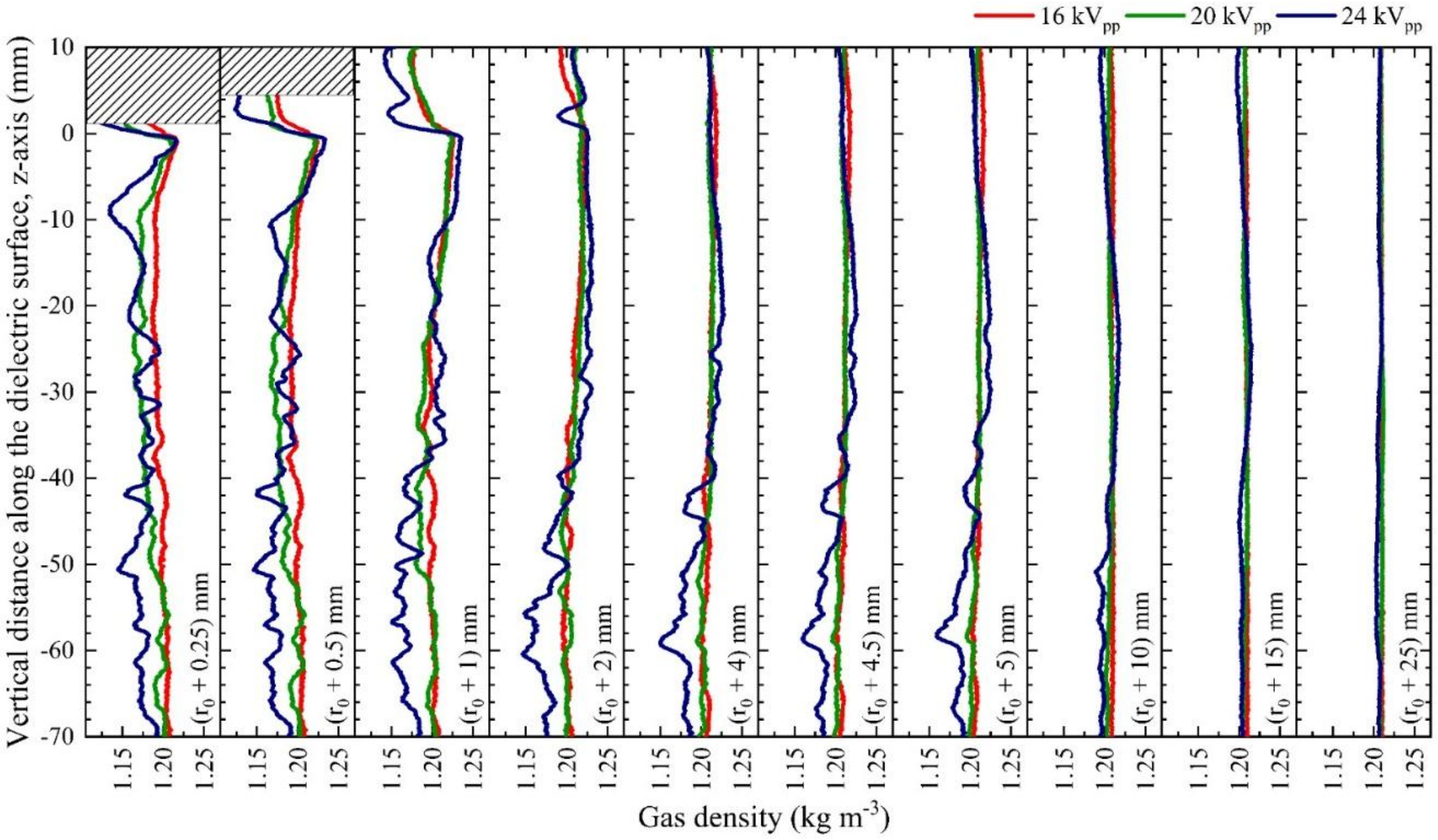

**Figure 8.** Gas density [Temperature (K) = 352.7 / Gas Density (kg m$^{-3}$)] along lines being parallel to the dielectric surface of the reactor and placed at various radial distances from the reactor symmetry axis ($r_0$ denotes the external radius of the reactor). The driving voltage is a variable parameter (16 $kV_{pp}$, 20 $kV_{pp}$, and 24 $kV_{pp}$). Ambient conditions: 291.5 K and 1.21 kg m$^{-3}$.

**Fig. 8** provides line plots with data from the absolute values of the air density, upstream and downstream of the driven electrode (*z*-axis), as one is moving far away from the dielectric surface (*i.e.*, radial distance; $r_0$ stands for the distance between the reactor symmetry axis and the external surface of the dielectric); the reader should keep in mind that the air temperature (K) equals the constant 352.7 divided by the gas density values of **Fig. 8**. In a 3D concept, these values refer to the surface points of different coaxial cylinders. At the radial distance $r_0$ + 0.25 mm and $r_0$ + 0.5 mm, values upstream of the driven electrode (positive *z*-axis) cannot be given inside the hatched rectangles, since they mark the electrode area itself. The line plots of **Fig. 8** mirror the abovementioned observations on the motif and the spatial variations of the air properties for increasing high voltage, while additional conclusions may be extracted. Namely:

- The influence of the present SDBD on the air, extinguishes at radial distances as long as about 15 mm from the dielectric surface, independently of the driving voltages tested here.
- The higher the externally applied electric field, the higher the air perturbation, without significant deviation of the local temperature from the ambient one. This fact implies electro-hydrodynamically rather than thermally induced flow fields.
- Within small radial distances (approx. < $r_0$ + 4 mm) the distinction between the laminar layer and turbulent domain is barely possible, especially at higher voltage levels.

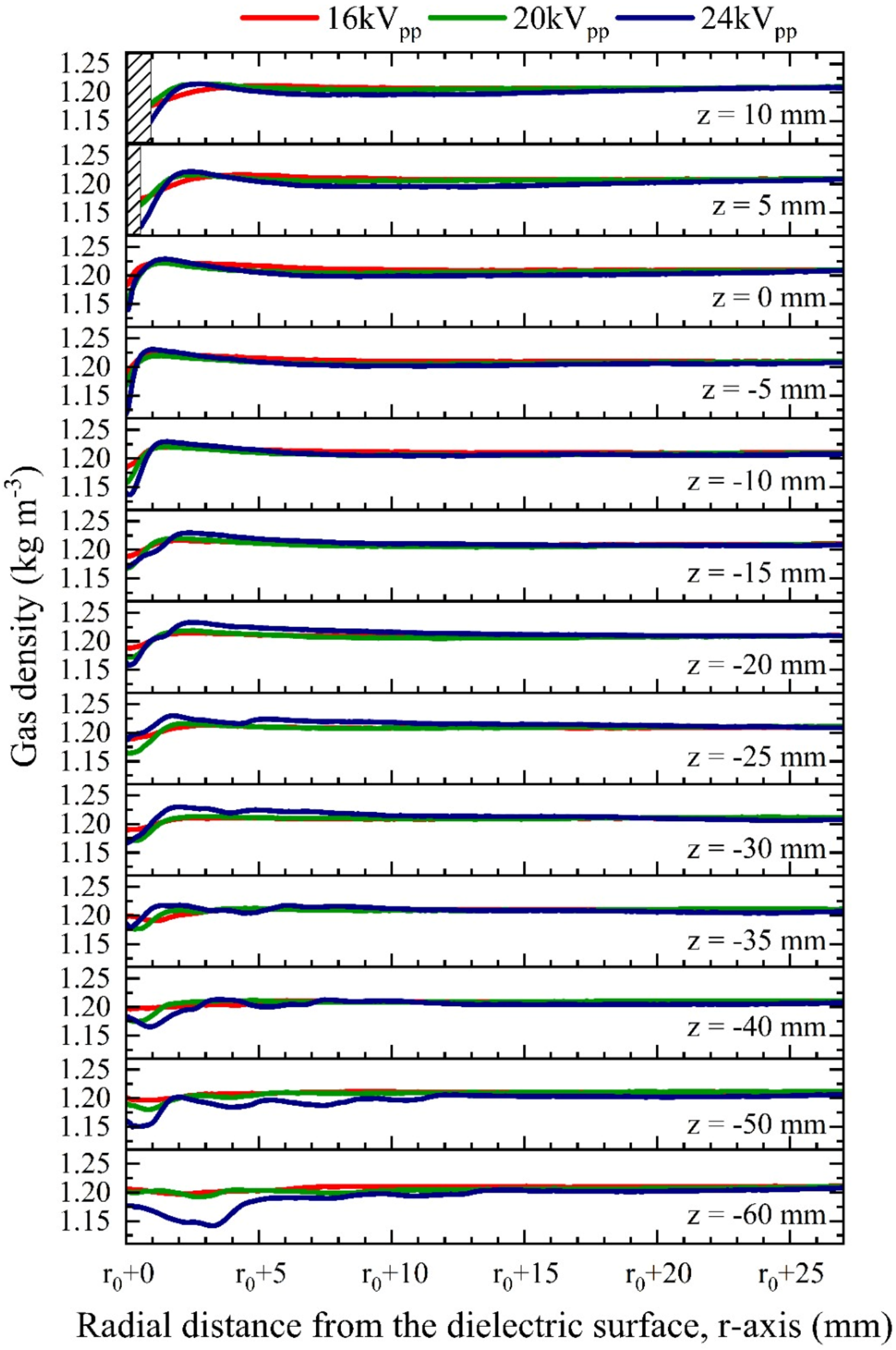

**Figure 9.** Gas density [Temperature (K) = 352.7 / Gas Density (kg $m^{-3}$)] as a function of the radial distance from the reactor symmetry axis ($r_0$ denotes the reactor external radius) at various vertical positions. The driving voltage is a variable parameter (16 $kV_{pp}$, 20 $kV_{pp}$, and 24 $kV_{pp}$). Ambient conditions: 291.5 K and 1.21 kg $m^{-3}$.

An alternative representation of the results can support the above statements and unveil additional useful information. Hence, on **Fig. 9** the acquired data on the air density are presented along the radial distance from the reactor symmetry axis at various vertical positions); the reader should keep in mind that the air temperature (K) equals the constant 352.7 divided by the gas density values of **Fig. 9**. In a 3D concept, these values refer to the surface points of different cylindrical disks stacked in parallel and having their centers on the reactor symmetry axis. The hatched rectangles in the first two frames correspond to the driven electrode, as in **Fig. 8**. After these plots, it is revealed that, at any level of the applied voltage, the air properties vary notably within small radial distances (approx. $< r_0 + 4$ mm) and they display a trend. Initially, within the glass/air interface the density has a dip, probably due to the increased plasma-induced velocity of the fluid-air. Then, for the next 3-5 mm, the density recovers and forms even a hump, most likely due to the mass continuity condition. Finally, for longer radial distances, the density (temperature) approaches the ambient one.

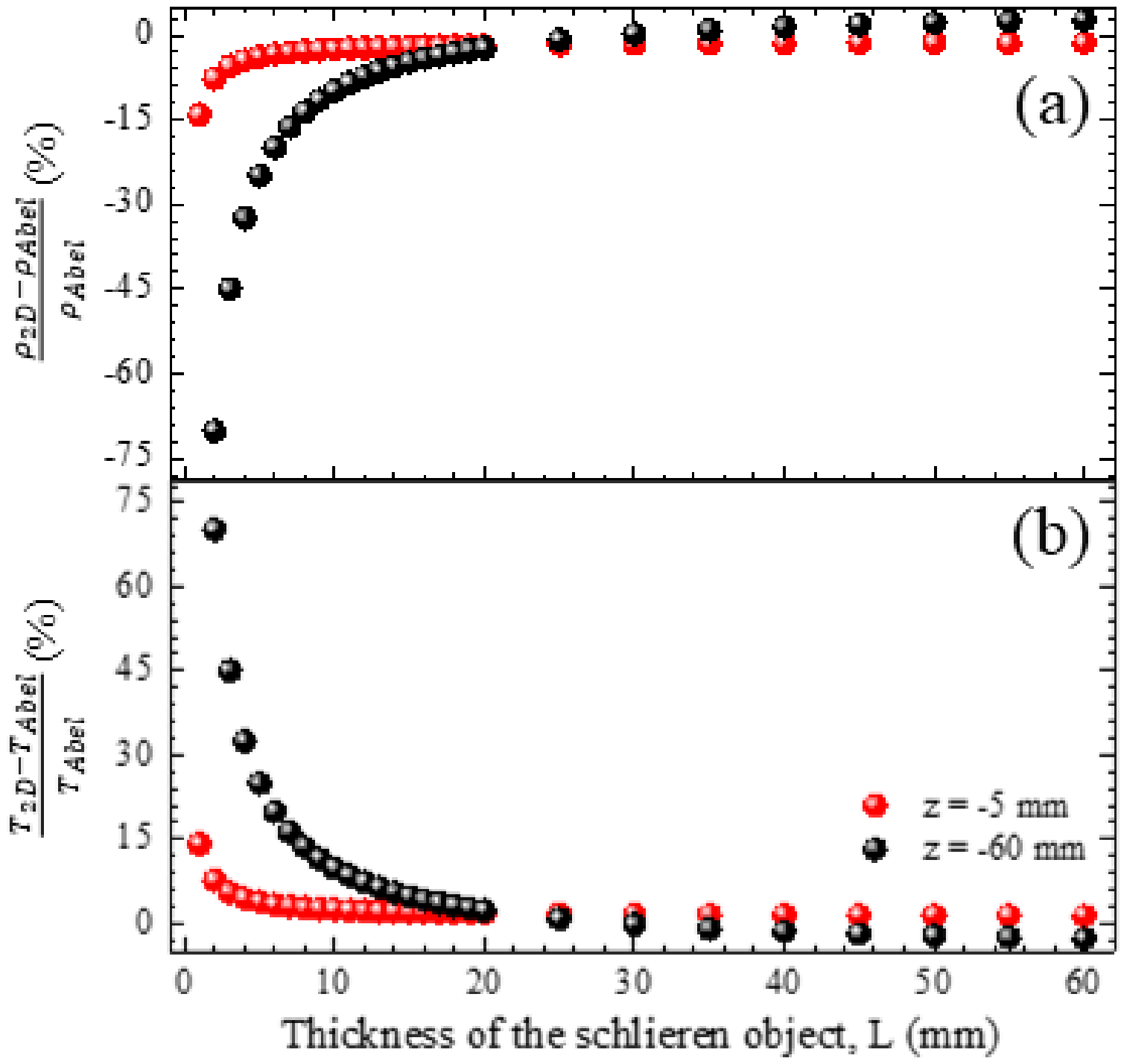

**Figure 10.** Relative (%) variation of the **(a)** air density and **(b)** temperature at two vertical positions (radial distance $r_0 + 2$ mm from the reactor symmetry axis) as a function of the thickness of the schlieren object, when Abel transformation is skipped. The relative values are given with respect to the corresponding values in the case that the cylindrical geometry of the SDBD reactor is considered and Abel transformation is applied.

Finally, in **Fig. 10**, two different cases are compared to demonstrate the erroneous conclusions that would be reached if Abel transformation was not implemented in the present work. In the first case, Abel transformation is not applied to the acquired image data and different thicknesses of the schlieren object are considered through the side-on observations. The cylindrical geometry of the present SDBD is thus ignored in this case, and a simplified probed area, a rectangle of width-depth $L$, being perpendicular to the line-of-sight of the observations, is assumed. In the second case, the cylindrical configuration is considered, and Abel transformation is applied as described in Section 3. **Fig. 10** illustrates the important variations that are introduced from a direct application of **Eq. 4** instead of **Eq. 6**. It is emphasized that the two plots represent the erroneous situation only for two points on the map of the flow field; ($r_0$ + 2 mm, –5 mm) and ($r_0$ + 2 mm, –60 mm). When the same comparison is continued for the entire map of the density and temperature fields, the resulting patterns are clearly dissenting.

Moreover, compared to literature, the gas density has been previously determined in AC-driven planar SDBDs using background-oriented Schlieren[54-56]. In general, the reported profiles and values are consistent to the present results. However, they cover only a relatively short region along the dielectric region (around 20-30 mm). Thus, the turbulent region has not been studied. Another point concerns the maximum air gas density. In ref. [**54]**, it is lower than that of the ambient gas within the entire gas medium, which contrasts with the present results (max ratio >1.024 at 10 kHz and 24 $kV_{pp}$) and other reported data; max. ratio 1.01 found for a planar configuration with silicone dielectric at 9.5 kHz and 12 $kV_{pp}$[55] and max. ratio 1.005 found for a planar configuration with polymethyl methacrylate dielectric at 1 kHz and 28 $kV_{pp}$[56].

The gas temperature has also been determined non-invasively through optical emission spectroscopy, but the studies have been limited either to space (*e.g.*, emissive area only) or time (*e.g.*, transient phases of the driving voltage only). Thus, for an AC-driven planar SDBD using epoxy printed circuit board as dielectric barrier, a mean value (over the space) of 380 K and 420 K has been measured at 1 kHz and 2 kHz, respectively[71]. Similarly, a flexible SDBD, having as dielectric barrier films of polyamide, has been investigated for sinusoidal driving voltage at 10 kHz[46]. The experiments gave a gas temperature of around 318 K at 12 $kV_{pp}$, 336 K at 14 $kV_{pp}$. and 410 K at 16 $kV_{pp}$. Moreover, differences in the gas temperature have been observed versus time, by means of optical emission spectroscopy. In a planar configuration with Plexiglas™ (PMMA; polymethyl methacrylate) dielectric (30 $kV_{pp}$ and 3 kHz), higher temperatures have been measured during the rising part of the voltage[72]. On the other hand, the opposite trend has been shown in a planar SDBD with quartz dielectric (24 $kV_{pp}$ and 8.4 kHz)[73]. Such effects can be further investigated and interpreted with the proper application of the present method in a time-resolved manner.

Overall, the variation of the gas density and temperature is due to the combined effect of the momentum transfer to neutrals and the heat development in SDBDs. As concerns the former, it is induced by the momentum transfer from the ions which are accelerated by electro-hydrodynamic forces[69]. For the latter, the heating sources may refer to the dielectric barrier and the surrounding gas[74-76]. The dielectric heating occurs primarily due to polarization effects, since the leakage current is practically negligible as far as good insulators are employed[74]. In the present work, the fused silica has resistivity as high as $10^{20}$ Ω m (at 20 °C), supporting this assumption. On the other hand, gas heating originates in the discharge kinetics; see electron (in)elastic collisions, rotational and vibrational excitation, ion–neutral collisions, *etc*[74]. Then, it has been stated that the final dielectric temperature is determined by convection (heat transfer between gas and dielectric) according to the induced air flow (on the order of 1 m $s^{-1}$[70])[76]. Finally, the distinguishment of the role of the electro-hydrodynamic forces and the heating in the flow field modulation remains a challenge, and additional tailored experiments and numerical modellings must be conducted.

## Conclusions

In this work, quantitative schlieren imaging along with inverse Abel transformation were properly applied and the plasma-induced flow field around the cylindrical surface of a dielectric barrier discharge was recorded at different operational conditions. The analyzed method led to gas density and temperature evaluation over the flow field patterns. The spatial profiles of those two parameters unveiled laminar and turbulent flow field regimes, the occurrence of sucking effect close to the driven electrode and suggested that electro-hydrodynamic force governed the flow field evolution. Maximum density and lowest temperature equal to 1.24 kg $m^{-3}$ and 285.4 K, respectively (position 0.84 mm from the dielectric surface, 0.59 mm downstream of the driven electrode edge), and lowest density and maximum temperature equal to 1.11 kg $m^{-3}$ and 318.8 K, respectively (position 0 mm from the dielectric surface, 6.97 mm downstream of the driven electrode edge), were identified with respect to the unperturbed air (1.21 kg $m^{-3}$ and 291.5 K, respectively). The degree of any perturbation was found to be dependent on the amplitude of the sinusoidal (10 kHz) high voltage which sustained the discharge.

## Acknowledgments

This work was funded by the Projects EIC “IgnitePLASMA” (G.A. No 101129853) and, partially, by the Project ANR ULTRAMAP (ANR–22–CE51–0027).

**Appendix**

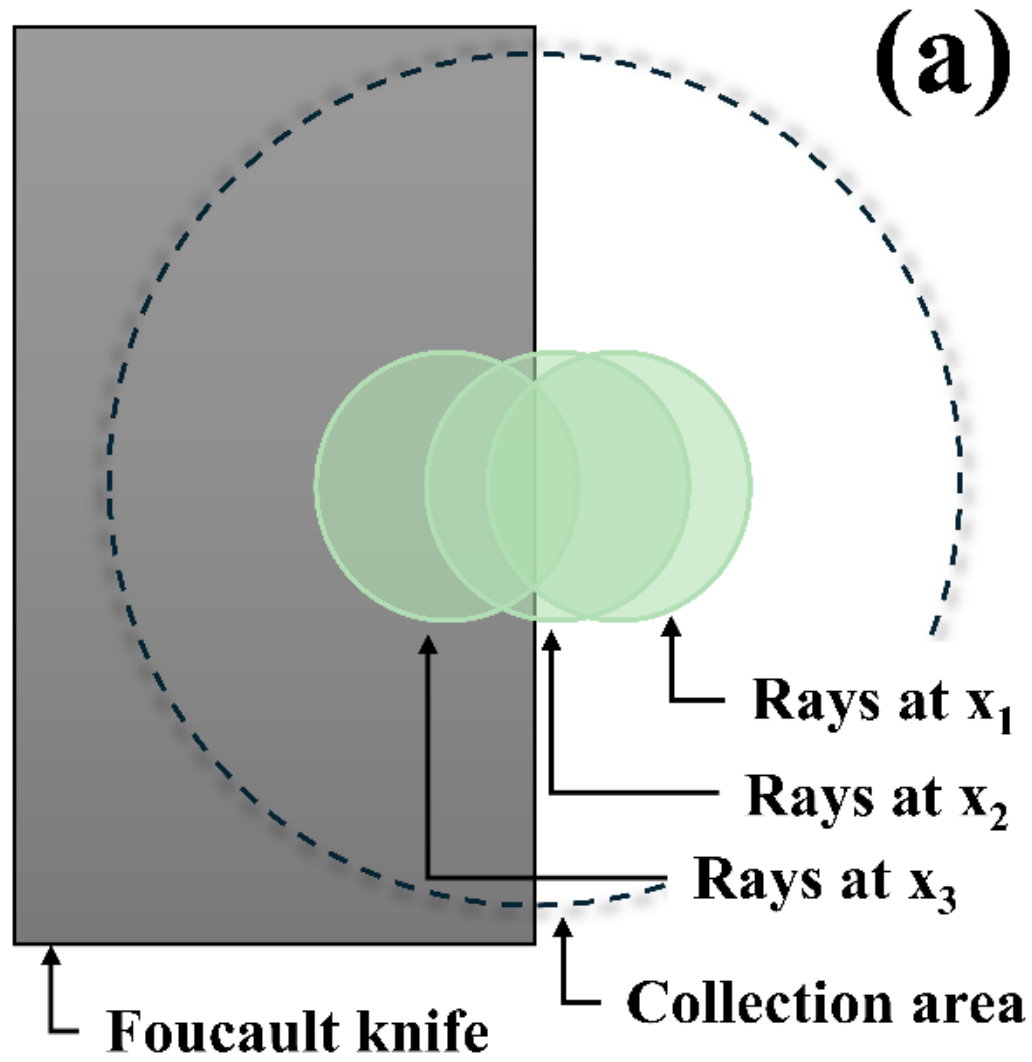

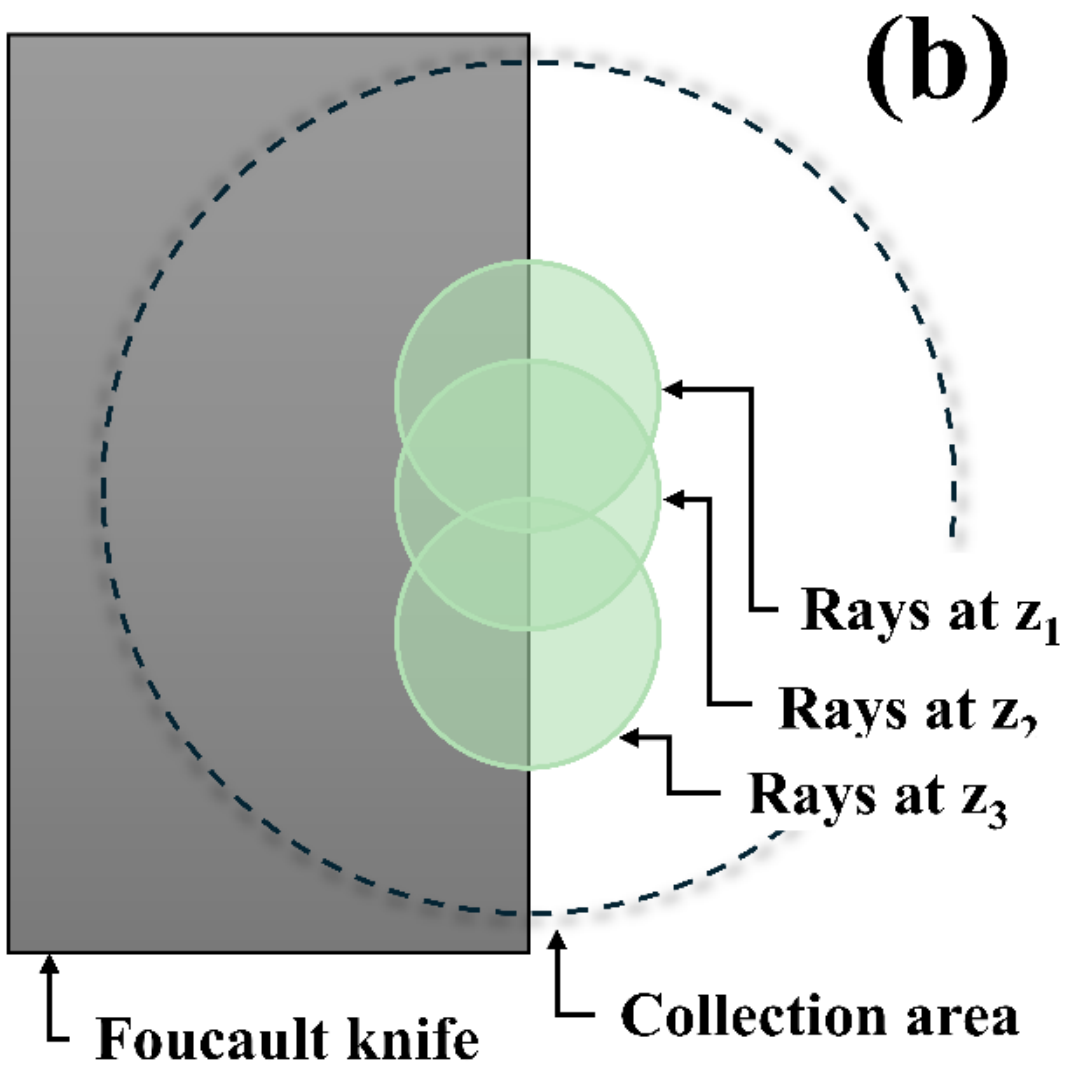

**Figure A1.** Conceptual representation of light rays deviated at three different random positions along **(a)** the *x*-axis only, **(b)** the *z*-axis only. The Foucault knife is here translated horizontally and thus the light intensity collected by the detector is affected in case (a) only.

Any ray deviation may be analyzed to a horizontal (**Fig. A1a**) and a vertical (**Fig. A1b**) displacement. Concerning the area of interest, *i.e.*, the light collection area, only horizontal displacements result in the variation of the light intensity recorded by the corresponding sensor (*i.e.*, camera in the case of present work). **Fig. A1a**. Any vertical displacement does not change the ray's portion which is masked by the Foucault knife (**Fig. A1b**).